%% file: rareDecays_Yb.tex
\RequirePackage{fix-cm}
\documentclass[twocolumn,epjc3]{svjour3}  

\smartqed  
\RequirePackage{graphicx}
\usepackage{hyperref}
\usepackage{color}

\usepackage{lineno}

\journalname{Eur. Phys. J. C}
\begin{document}
\include{commands}
\title{Search for rare alpha and double beta decays of Yb isotopes to excited levels of daughter nuclei}


\author{M. Laubenstein\thanksref{e1,addr1}
        \and
        B. Lehnert\thanksref{e2,addr2} 
        \and
       S. S. Nagorny\thanksref{e3,addr3} 
        \and
       S. Sch\"{o}nert\thanksref{e4,addr4} 
}

\thankstext{e1}{e-mail: matthias.laubenstein@lngs.infn.it}
\thankstext{e2}{e-mail: bjoernlehnert@lbl.gov}
\thankstext{e3}{e-mail: sn65@queensu.ca}
\thankstext{e4}{e-mail: schoenert@ph.tum.de}


\institute{INFN - Laboratori Nazionali del Gran Sasso, 67100 Assergi (AQ), Italy \label{addr1}
           \and
           Nuclear Science Division, Lawrence Berkeley National Laboratory, Berkeley, CA 94720, U.S.A. \label{addr2}
           \and
           Queen's University, Physics Department, Kingston, ON, K7L 3N6, Canada \label{addr3}
           \and
           Physik Department, Technische Universit\"{a}t M\"{u}nchen, Germany \label{addr4}
}

\date{Received: date / Accepted: date}

%
%
%
%
%

\maketitle

\begin{abstract}
A search for alpha and double beta decays of ytterbium isotopes was performed with an ultra-low background high purity germanium detector at Gran Sasso Underground Laboratory (Italy). A 194.7~g Yb$_2$(C$_2$O$_4$)$_3$ powder sample was measured for 11.3~d with a total Yb exposure of 1.25~kg$\times$d. 
Half-life limits for $\alpha$-decay modes of \nuc{Yb}{168}, \nuc{Yb}{170}, \nuc{Yb}{171}, \nuc{Yb}{172}, \nuc{Yb}{173},  \nuc{Yb}{174} and \nuc{Yb}{176} into the first excited states have been obtained between \baseT{6}{14}~yr and \baseT{2}{16}~yr. 
These are the first experimental constraints of these decay modes. 
Double electron capture of \nuc{Yb}{168} and double beta decay of \nuc{Yb}{176} into the first excited 2$^+$ and 0$^+$ states could be excluded with limits between \baseT{1}{14}~yr to \baseT{8}{16}~yr. This improves the experimental information on some of the decay modes compared to previous constraints.

\end{abstract}

\section{Introduction}
\label{intro}

The study of rare nuclear decays is an active field of research with steady improvements utilizing more and more sensitive detectors. Experimental information on rare decays helps understanding nuclear structure, has applications in a variety of fields as nuclear chronometers and is relevant as long-lived backgrounds in other rare events searches. In the case of neutrinoless double beta decay it can even shed light on fundamental concepts beyond the Standard Model of particle physics. 
Those searches are often targeted to one particular purpose; however, in this work we pursued a more generic approach, investigating many different rare alpha and double beta decays in ytterbium isotopes.

Typically, two fundamentally different detection approaches are used. One where the target isotope is embedded in the detector material (``source$\, =\, $detector'') and one where the target isotope is external to the detector (``source$\, \neq\, $detector''). 
Recent examples for the first method are searches for \nuc{W}{180} $\alpha$-decay with the isotopes embedded in scintillating crystals: 
\nuc{CdWO_4}{116} \cite{Danevich03,Cozzini04}, 
CaWO$_4$ \cite{Zdesenko05,Zdesenko05_v2}, and 
ZnWO$_4$ \cite{Belli11}. 
Limits for $\alpha$-decays in lead isotopes were set with PbWO$_4$ detectors \cite{Beeman13}
and $\alpha$-decays in \nuc{Eu}{151} were investigated with CaF$_2$(Eu) crystals \cite{Belli07}. 
A particularly famous example is \nuc{Bi}{209} which was long thought to be the heaviest stable isotope. In 2003, the $\alpha$-decay of \nuc{Bi}{209} was discovered with a half-life of \baseT{1.9}{19}~yr \cite{Marcillac03}, the longest $\alpha$-decay half-life ever measured. This removed bismuth from the list of elements with at least one stable isotope. In 2011 also the \nuc{Bi}{209} $\alpha$-decay transition to the first excited state was observed \cite{Beeman11}. Both experiments were performed with scintillating bolometers based on Bi$_4$Ge$_3$O$_{12}$ (BGO) detectors. 
A major advantage of the ``source$\, =\, $detector'' approach is almost 100\% detection efficiency as well being able to detect alpha's and beta's directly inside bulky large-mass target. The disadvantage is the limitation to certain target elements / isotopes which are suitable for detector manufacturing.  

In contrast, the ``source$\, \neq\, $detector'' approach allows to measure virtually all isotopes of interest with well established and high performing detector systems e.g.\ ionisation chambers or high purity germaniums (HPGe) detectors.
Ionization chambers for alpha-spectroscopy allow the direct measurement of alpha particles \cite{alphaChamber1} with recent examples of half-life measurements in \nuc{Sm}{147} \cite{Sm147} and \nuc{Pt}{190} \cite{Pt190}. Precision half-life measurements in the \baseTsolo{11}~yr range are achieved but ultimately the sensitivity is limited since the sample thickness cannot be larger than a few microns.  
With HPGe \gray-spectroscopy the detection is limited to decay modes containing \grays\ in the final state. 
%
Recent examples for searches of $\alpha$-decays to excited levels of the daughter nucleus with HPGe \gray-spectroscopy are 
in dysprosium (using a Dy$_2$O$_3$ powder sample \cite{Belli11_2}), 
in europium (Eu$_2$O$_3$ powder \cite{Danevich12}), 
in platinum (Pt metal \cite{Belli11_3}) and 
in osmium (Os metal \cite{Belli13}) isotopes. 
The best achieved sensitivities are on the level of $T_{1/2} > 10^{18-20}$~yr using low-background HPGe detectors in underground laboratories, optimizing the sample-detector geometry and using purified samples. For a recent review see \cite{alphaReview}.

In this work we extend the experimental information on rare $\alpha$-decays in Yb isotopes using an ultra low background HPGe detector setup in the ``source$\, \neq\, $detector'' configuration. 
Ytterbium contains 7 quasi-stable isotopes (\nuc{Yb}{168}, \nuc{Yb}{170}, \nuc{Yb}{171}, \nuc{Yb}{172}, \nuc{Yb}{173},  \nuc{Yb}{174} and \nuc{Yb}{176}) which can undergo $\alpha$-decay.

In addition, \nuc{Yb}{176} can undergo double beta decay. The two-neutrino mode (\bb{2}) is a second order standard model process with half-lives in the range of \baseTsolo{18-21}~yr. It has been observed in 11 isotopes using the ``source$\, =\, $detector'' approach. Decays into excited states of the daughter isotope have been observed in \nuc{Nd}{150} and \nuc{Mo}{100} \cite{ESAverage} using the ``source$\, \neq\, $detector'' approach and HPGe detectors. The neutrinoless mode (\bb{0}) is subject of intensive research with experiments reaching the tonne-scale in target mass. It has not been observed and would imply lepton number violation, the Majorana nature of neutrinos and help constraining the absolute neutrino mass scale \cite{DBDreview}.

The opposite processes of double electron capture (\bbe{2} and \bbe{0}) can occur in \nuc{Yb}{168}. The first compelling evidence of \bbe{2} has recently been observed in \nuc{Xe}{124} with a half-life of \baseT{1.8}{22}~yr \cite{apr19} making it the slowest nuclear decay ever observed. 
Special cases are \bbe{0} modes which nominally have no final state particles other than two x-rays from the two captured electrons. The remaining energy can be released by a Bremsstrahlung photon, carrying the remaining decay energy \cite{Doi93}. 
Other allowed processes are two Bremsstrahlung photons or an $e^-$-$e^+$ pair in case of sufficiently available energy. However, the latter processes have more complex experimental signatures and are also suppressed by additional vertices. Hence this work focuses on the decay with one Bremsstrahlung photon. The two captures can occur from different electron shells with binding energies $E_{\epsilon 1}$ and $E_{\epsilon 2}$, removing energy from the Bremsstrahlung photon $E_{\gamma}$. The available energy is
$E_{\gamma} = Q - E_{\rm level} - E_{\epsilon 1} - E_{\epsilon 2} $
%
with respect to the Q-value and the energy of the nuclear final state $E_{\rm level}$. 
K-shell and L-shell binding energies are about 57.5~keV and 9.7~keV, respectively in erbium \cite{NuclDataX}. Variations inside the shells are largely within the experimental energy resolution and are neglected. KK-shell captures are most likely, taking a total of 115.0~keV in binding energy. However, KK-shell captures are not allowed for decays into $0^+$ states of the daughter nucleus \cite{Doi93}  making KL-shell captures with 67.2~keV binding energy the most likely ones for these cases. Only the most likely capture combinations are considered in this work. For a recent review on neutrinoless double electron capture see \cite{0neeReview}.

\begin{table*}[t]
\begin{center}
\begin{tabular}{llllllll}
\hline
isotope &  abundance & daughter & Q-value & mode &  level   & T$_{1/2}^{\rm th}$ & T$_{1/2}^{\rm exp}$ previous   \\
             &        [\%]          &            &  [keV]    &            & $J^\pi$ [keV]   &     [yr]                    &  [yr]             \\
\hline

\nuc{Yb}{168} & 0.123(3)   &  \nuc{Er}{164} & 1936.1(1.2)   &  $\alpha$  &  $2^+_1$ 91.4 &      \baseT{4.4}{24} - \baseT{1.1}{27}  &   -               \\
		      &                  &  \nuc{Er}{168} &  1409.3(1.7)  &  $2\nu\epsilon\epsilon$  &  $2^+_1$ 79.8     &   -  &   \baseT{>2.3}{15}  \cite{Belli19_Yb168}             \\
		      &                  &  			  & 			  & $2\nu\epsilon\epsilon$  &  $2^+_2$ 821.2   &    - &   \baseT{>4.4}{17}   \cite{Belli19_Yb168}            \\
		      &                  & 			  & 			  & $2\nu\epsilon\epsilon$  &  $0^+_1$ 1217.2  &    \baseT{5.4}{33} \cite{Ceron99}   &   \baseT{>1.5}{18}   \cite{Belli19_Yb168}            \\
		      &                  & 			  & 			  & $0\nu\rm KL$  &  $0^+_{gs}$ 0  &    -   &   \baseT{>6.9}{17}  \cite{Belli19_Yb168}              \\
		      &                  & 			  & 			  & $0\nu\rm KK$  &  $2^+_1$ 79.8  &    -    &   \baseT{>4.4}{14}  \cite{Belli19_Yb168}             \\
		      &                  & 			  & 			  & $0\nu\rm KK$  &  $2^+_2$ 821.2   &  -      &   \baseT{>3.9}{17}  \cite{Belli19_Yb168}              \\
		      &                  & 			  & 			  & $0\nu\rm KL$  &  $0^+_1$ 1217.2  &  -    &   \baseT{>1.5}{18}   \cite{Belli19_Yb168}             \\

\nuc{Yb}{170} & 2.982(39) &  \nuc{Er}{166}  &    1737.2(1.2)  & $\alpha$  &  $2^+_1$ 80.6    &   \baseT{5.1}{31}- \baseT{2.1}{34}  &  -               \\

\nuc{Yb}{171} & 14.09(14) & \nuc{Er}{167}   &    1559.5(1.2) &  $\alpha$  &  $9/2^+_1$ 79.3    &   \baseT{2.5}{38}- \baseT{5.5}{39}  &   -              \\

\nuc{Yb}{172} & 21.68(13) & \nuc{Er}{168}   &    1310.8(1.2)  &  $\alpha$  &  $2^+_1$ 79.8     &     &   -              \\

\nuc{Yb}{173} & 16.103(63) & \nuc{Er}{169}   &    947.0(1.2)  &  $\alpha$  &  $3/2^+_1$ 64.6   &     &   -              \\

\nuc{Yb}{174} & 32.026(80) & \nuc{Er}{170}   &    739.3(1.3) &  $\alpha$  &  $2^+_1$ 78.6    &     &   -              \\

\nuc{Yb}{176} & 12.996(83) & \nuc{Er}{172}  &  567(4)           &  $\alpha$  &  $2^+_1$ 77.0 &      \baseT{7.2}{95}- \baseT{1.1}{100}  &   -              \\
                      &                     & \nuc{Hf}{176}  &  1085.0(1.5)   & $2(0)\nu\beta\beta$  &  $2^+_1$ 88.3 &  -    &   \baseT{>4.5}{16}  \cite{Belli19_Yb168} (*)             \\

\hline
\end{tabular}
\medskip
\caption{\label{tab:isotopes} Isotopes and decay modes investigated in this work. Shown is the Yb isotope, the natural isotopic abundance, the daughter isotope, the decay mode, the level state and energy, the Q-value, the theoretical half-life as discussed in the text and previous experimental constraints. Nuclear data were taken from \cite{NuclData}.
(*) Note that half-lives in \cite{Belli19_Yb168} are given for \bb{2} and \bb{0} modes separately. We consider the experimental signature to be identical for both modes since the $\beta$'s will not contribute to the detector signal.
}
\end{center}
\end{table*}

\begin{figure*}
  \centering
  \includegraphics[width=0.99\textwidth]{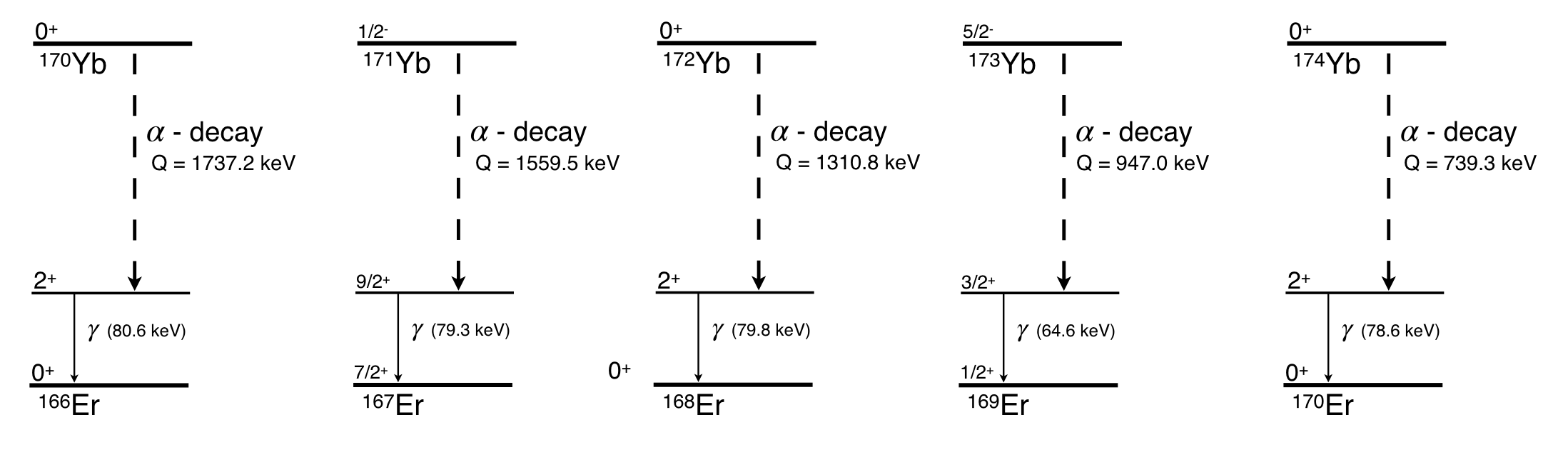}
\caption{Decay schemes of Yb isotopes with $\alpha$-decay modes as investigated in this work. Data taken from \cite{NuclData}.}
  \label{pic:decayScheme}
\end{figure*}

\begin{figure*}
  \centering
  \includegraphics[width=0.99\textwidth]{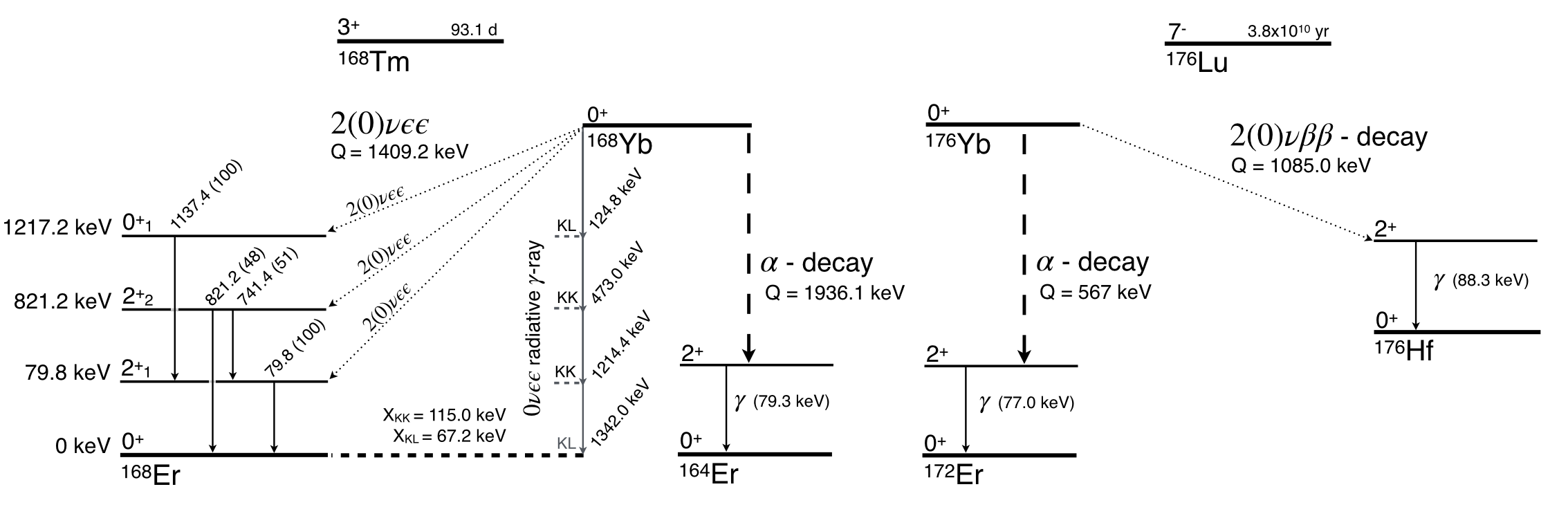}
\caption{Decay schemes of Yb isotopes with double beta decay and $\alpha$-decay modes as investigated in this work. \nuc{Yb}{168} has multiple $\epsilon\epsilon$ decay modes: \bbe{2} is only detectable if the decay goes to one of the excited states. \bbe{0} could also occur radiatively, adding a Bremsstrahlung \gray\ to the final state. The \gray\ energy depends on the atomic shells of two electron captures, sharing the available energy with a combination of $\rm X_{KK}$ or $\rm X_{KL}$. Angular momentum conservation does not allow the KK capture combination for decays into $0^+$ states. 
Data taken from \cite{NuclData}. }
  \label{pic:decayScheme2}
\end{figure*}

All investigated Yb isotopes and decay modes are listed in \tab \ref{tab:isotopes}. The natural abundances of the isotopes, the decay daughters, the Q-values, and the investigated excited levels are listed. Decay schemes for all investigated Yb isotopes and transitions are shown in \fig \ref{pic:decayScheme} and \ref{pic:decayScheme2}.
 
All decay modes require a \gray\ emissions as experimental signature to be detected in the ``source$\, \neq\, $detector'' approach. For $\alpha$-decays, the half-life scales with the available energy by $\log{T_{1/2}} \propto E^{-1/2}$ according to the Geiger-Nuttall law \cite{GeigerNuttall}. We thus limit the search to the lowest excited states for $\alpha$-decays which are in the range of 60 to 100~keV for all Yb isotopes. 
For the $\beta\beta$ and $\epsilon\epsilon$ modes the half-life scales with the available energy by $T_{1/2}\propto E^{-11}$ and $T_{1/2} \propto E^{-5}$ for the $2\nu$ and $0\nu$ modes, respectively. Since all double beta decay isotopes have a $0^+$ ground state, the most likely transitions are to a $0^+$ states of the daughter. Spin suppression occurs for larger angular momentum transfers. Hence, we limit the search for double beta decays to the first excited $0^+$ state and all lower lying $2^+$ states. 

Calculated half-life estimates for the $\alpha$-decays are based on the cluster models \cite{Buck91,Buck92}, phenomenological fission theory of $\alpha$-decay \cite{Poenaru83} and semi-classical WKB approximation \cite{Brown92}.
Experimental limits on $\alpha$-decays to the ground state exist only from a broad search in the 1950's \cite{alphaYb168} with an experimental half-life sensitivity of \baseTsolo{17}~yr assuming 100\% isotopic abundance\footnote{This generic experimental sensitivity has to be multiplied with the isotopic abundance of the isotope.}. This work gives the first constraints on $\alpha$-decays into excited states for all Yb isotopes.
 
Predictions for double beta decay half-lives are more difficult, requiring a nuclear matrix element obtained in a specific nuclear model framework.
Predictions for excited state transitions are less frequently found in literature compared to ground state transitions. The only data is available for the \nuc{Yb}{168} \bbe{2} $0^+_1$ transition at \baseT{5.4}{33}~yr (corresponding ground state transition at \baseT{2.0}{23}~yr) \cite{Ceron99}. For \nuc{Yb}{176},  ground state transitions are predicted in \cite{Delion17} ($2\nu$: \baseT{1.3}{26}~yr) and in \cite{Hirsch02} ($2\nu$: \baseT{2.8}{23}~yr, $0\nu$ at $m_{\beta\beta}=1$~eV: \baseT{6.4}{25}~yr).
Previous limits on double beta decay modes in \nuc{Yb}{168} and \nuc{Yb}{176} were established in \cite{Belli19_Yb168}, and are listed in the last column of \tab \ref{tab:isotopes}.

\section{Experimental Setup and Sample}

The experimental setup used for this work is the $\gamma$-spectrometry HPGe detector ``GeMPI” \cite{gempi1,gempi2}, one of the world’s most sensitive low-background detectors. 
It is located underground in the Gran Sasso National Laboratories of the I.N.F.N. (Italy) which provides an average shielding of 3600~m~w.e.\ overburden against cosmic muons.
The coaxial p-type germanium detector has 405 cm$^3$ active volume and 2.15~kg mass, corresponding to 100.0\% efficiency relatively to a 3''$\times$3'' NaI(Tl) detector. 
The energy resolution of the spectrometer is 2.0~keV at the 1332~keV \nuc{Co}{60} \gline. 
%
The detector is shielded by 20~cm of low-radioactivity lead with $<133$~Bq/kg in the outer layer, 32~Bq/kg in a central layer and $< 5$~Bq/kg in the inner layer. 
Inside the lead shield is a 5~cm copper layer. 
In order to remove radon, the setup is continuously flushed with nitrogen which stems from the gas phase of a liquid nitrogen storage tank. 
Interlocks for sample insertion and Rn daughter suppression are all enclosed in an air tight aluminum and steel housing (1~mm thick) with butyl rubber gloves (0.8~mm). 
The detector has a big sample chamber of $250\times250\times220$~mm.

Various samples containing Ytterbium have been investigated:
Yb$_2$O$_3$ powder, Yb(NO$_3$)$_3$ solution, and Yb$_2$(C$_2$O$_4$)$_3$ powder. The latter was found to be best suited due to its low internal background which is shown in \tab \ref{tab:bgContent}. The activity estimates are from $\gamma$-spectroscopy and MaGe Monte Carlo simulations \cite{bos11} based on Geant4. Limits are set with the Feldman Cousins method \cite{Feldman98}.
No evidence for background from the natural decay chains \nuc{U}{235}, \nuc{U}{238}, and \nuc{Th}{232} are observed in the sample. Upper limits are set on the mBq/kg level. Other common radionuclides, such as \nuc{K}{40} from natural radioactivity or \nuc{Cs}{137} from artificial origins, are also not observed. 
However, a significant activity of \nuc{Yb}{169}, \nuc{Yb}{175}, and \nuc{Lu}{176} was measured.
Short-lived \nuc{Yb}{169} (32.0~d) and \nuc{Yb}{175} (4.2 d) isotopes are produced by cosmic rays via neutron spallation and through thermal neutron capture on ytterbium isotopes present in its natural abundance. The relatively high observed activities of 22 and 150~mBq/kg for \nuc{Yb}{169} and \nuc{Yb}{175}, respectively, are due to sample storage above ground, followed by the underground measurement shortly afterwards.
%
%
The presence of \nuc{Lu}{176} (2.6\% natural abundance in lutetium, \baseT{T_{1/2}=3.76}{10}~yr) is expected due to similar chemical properties of lutetium and ytterbium which leads to difficulties in chemical separation.

\begin{table}[htbp]
\begin{center}
\begin{tabular}{ll}
\hline
 nuclide  &  activity [mBq/kg]     \\
\hline
  \nuc{Ra}{226}   & $< 5.0$    \\ 
  \nuc{Pa}{234m}   & $<140$    \\ 
  \nuc{U}{235}   & $< 4.0$    \\ 
  \nuc{Ra}{228}   & $< 4.0$   \\ 
  \nuc{Th}{228}   & $< 5.0$    \\ 

  \nuc{K}{40}   & $< 17$    \\ 
  \nuc{Co}{60}   & $< 2$  \\ 
  \nuc{Cs}{137}   & $< 2$    \\ 
  \nuc{Yb}{169}   & $22\pm5$   \\ 
  \nuc{Yb}{175}   & $150\pm36$    \\ 

  \nuc{Lu}{176}   & $15\pm2$   \\ 

\hline
\end{tabular}
\medskip
\caption{\label{tab:bgContent} Radioactive contamination of the 194.7~g Yb$_2$(C$_2$O$_4$)$_3$ powder sample. The upper limits are given at 90\% C.L., and the uncertainties of the measured activities at 68\% C.L.
}
\end{center}
\end{table}

ICP-MS measurements for complementary background estimates were not possible for this sample. The isotopic abundances of Yb isotopes are taken as the natural abundance from literature \cite{NuclData}.

\begin{figure*}
  \centering
  \includegraphics[width=0.99\textwidth]{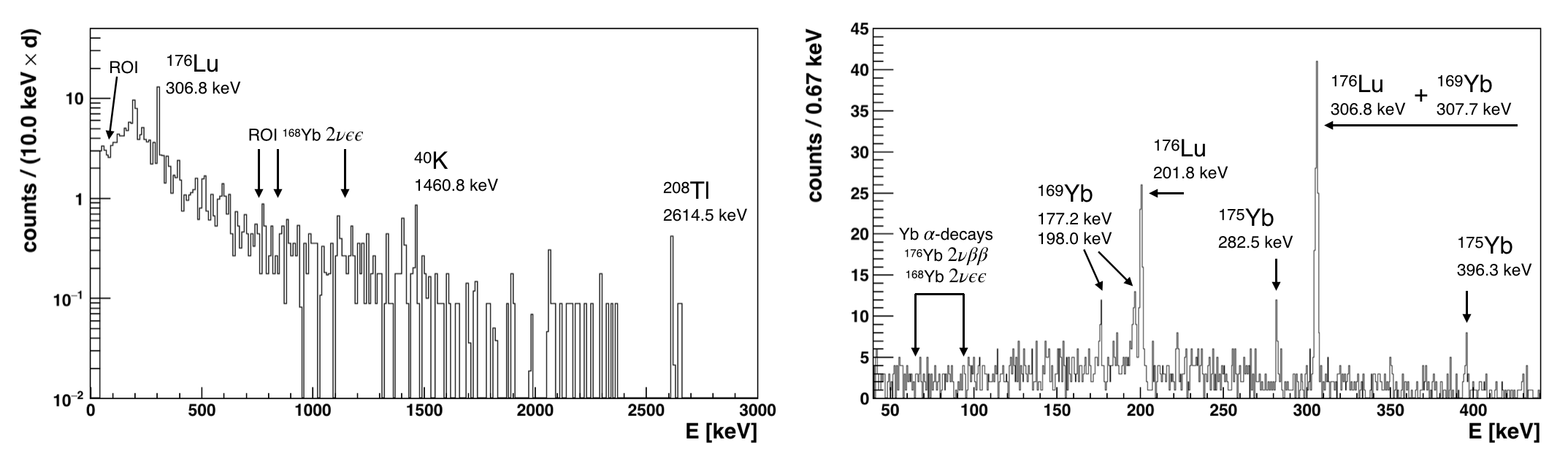}

\caption{Measured HPGe spectrum of the 194.7~g Yb$_2$(C$_2$O$_4$)$_3$ powder sample obtained in 11.3 days. Left: full spectrum normalized to counts per day. Right: zoom in to the low energy region in original DAQ binning. Prominent background peaks and regions of interest are highlighted. }
  \label{pic:WideSpec}
\end{figure*}

For the measurement, 194.7 g of Yb$_2$(C$_2$O$_4$)$_3$ powder was contained in a plastic container and placed directly on the endcap of the ultra-low background HPGe detector. 
Data was taken for 11.3~days. The full spectrum is shown in \fig \ref{pic:WideSpec} on the left and a zoom into the low energy region is shown on the right. Prominent peaks and the regions of interests are labeled.

\section{Analysis}

The analysis is a peak search for the de-excitation \grays\ from each decay mode independently. 
The peak fits are performed in a Bayesian framework using the Bayesian Analysis Toolkit (BAT) \cite{Caldwell:2009kh}. 
The likelihood $\mathcal{L}$ is defined as the product of the Poisson probabilities over each bin $i$ for observing $n_{i}$ events while expecting $\lambda_{i}$ events. $\lambda_{i}$ is the sum of the signal $S_i$ and background $B_i$ expectation: 

\begin{eqnarray}
\label{eq:Likelihood}
\mathcal{L}(\mathbf{p}|\mathbf{n}) =
 \prod  \limits_i \frac{\lambda_{i}(\mathbf{p})^{n_{i}}}{n_{i}!} e^{-\lambda_{i}(\mathbf{p})}\ \ ,\ \  \lambda_{i}(\mathbf{p}) = S_i + B_i\ .
\end{eqnarray}

Here \textbf{n} denotes the data and \textbf{p} the set of floating parameters.

$S_i$ is the integral of the Gaussian peak shape in bin $i$ given the total signal peak counts $s$
\begin{eqnarray}
S_i&=&
 \int_{\Delta E_{i}}  \frac{s}{\sqrt{2\pi}\sigma_E} 
\cdot \exp{\left(-\frac{(E-E_0)^2}{2\sigma_E^2}\right)} dE\ , \label{eq:Si}
\end{eqnarray}

where $\Delta E_{i}$ is the bin width, $\sigma_E$ the energy resolution, and $E_0$ the \gline\ energy as the mean of the Gaussian.
$B_i$ is the background expectation 
\begin{eqnarray}
B_i = \int_{\Delta E_{i}}&& b + c\left( E-E_0 \right) \\
+&&   \sum \limits_{l} \left[\frac{b_{l}}{\sqrt{2\pi}\sigma_{l}} \cdot \exp{\left(-\frac{(E-E_{l})^2}{2\sigma_{l}^2}\right)}\right] dE \nonumber
\label{eq:Bi}
\end{eqnarray}

which is implemented as a linear function (parameters $b$ and $c$) and $l$ Gaussian background peaks in the fit window, depending on the decay mode.

The signal counts are connected with the half-life $T_{1/2}$ of the decay mode as
\begin{eqnarray}
\label{eq:HLtoCounts}
s =
\ln{2} \cdot  \frac{1}{T_{1/2}} \cdot \epsilon \cdot N_A \cdot T \cdot m \cdot f  \cdot \frac{1}{M}\ ,
\end{eqnarray}

where $\epsilon$ is the full energy peak detection efficiency,
$N_A$ is Avogadro's constant,
$T$ is the live-time (11.3~d), 
$m$ is the mass of Yb in the sample (110.4~g), 
$f$ is the isotopic fraction of the respective Yb isotope, 
and $M$ is the molar mass of natural Yb (173.05).

\begin{table*}[htbp]
\begin{center}
\begin{tabular}{llllll}
\hline
nuclide (decay) & daughter (level)  &    \glines\ energies  &  $\epsilon$   & $\sigma_{\rm res}$ &  T$_{1/2}$ (90\% C.I.) \\
                         &  ($J^\pi $ keV)  &     [keV]                    &  [\%]               &  [keV]                      &    [yr] \\
\hline

\nuc{Yb}{168} ($\alpha$) & \nuc{Er}{164} ($2^+_1 91.4)$   &   91.4    & 0.026  &   0.80  &   \baseT{>6.3}{14}               \\
\nuc{Yb}{170} ($\alpha$) & \nuc{Er}{166} ($2^+_1 80.6)$   &   80.6    & 0.0060   &   0.80  &   \baseT{>2.2}{15}          \\
\nuc{Yb}{171} ($\alpha$) & \nuc{Er}{167} ($9/2^+_1 79.3)$   &   79.3    & 0.0058 &   0.79  &   \baseT{>1.1}{16}         \\
\nuc{Yb}{172} ($\alpha$) & \nuc{Er}{168} ($2^+_1 79.8)$   &   79.8    &   0.0059   &    0.80 &    \baseT{>1.8}{16}       \\
\nuc{Yb}{173} ($\alpha$) & \nuc{Er}{169} ($3/2^-_1 64.6)$   &   64.6    &  0.0002    & 0.79  &    \baseT{>7.8}{14} (*)       \\
\nuc{Yb}{174} ($\alpha$) & \nuc{Er}{170} ($2^+_1 78.6)$   &   78.6    &   0.0043    & 0.79 &    \baseT{>1.8}{16}         \\
\nuc{Yb}{176} ($\alpha$) & \nuc{Er}{172} ($2^+_1 77.0)$   &   77.0    & 0.0033   &   0.79  &   \baseT{>4.0}{15}          \\
\hline
\nuc{Yb}{168} (\bbe{2}) & \nuc{Er}{168} ($2^+_1 79.8)$   &   79.8    & 0.0057   &   0.80  &   \baseT{>1.0}{14}               \\
\nuc{Yb}{168} (\bbe{2}) & \nuc{Er}{168} ($2^+_2 821.2)$   &   741.4    & 1.12   &   0.95  &      \baseT{>2.4}{16}             \\
											 &  &   821.2    & 1.01   &   0.97  &            \\
\nuc{Yb}{168} (\bbe{2}) & \nuc{Er}{168} ($0^+_1 1217.2)$   &   1137.4    & 1.81   &   1.03  &   \baseT{>7.8}{16}               \\
\nuc{Yb}{168} ($0\nu\rm KL$) & \nuc{Er}{168} ($0^+_{gs}\, 0)$   &   1342.0   & 1.66   &   1.06  &    \baseT{>4.7}{16}               \\
\nuc{Yb}{168} ($0\nu\rm KK$) & \nuc{Er}{168} ($2^+_1 79.8)$   &   79.8    & 0.0042   &   0.80  &    \baseT{>5.4}{16}                 \\
											&  &   1214.4    & 1.76   &   1.04  &            \\
\nuc{Yb}{168} ($0\nu\rm KK$) & \nuc{Er}{168} ($2^+_2 821.2)$   &   741.4    & 0.99   &   0.95  &     \baseT{>3.0}{16}            \\
											 &  &   821.2    & 0.87   &   0.97  &            \\
											 &  &   473.0    & 2.86   &   0.90  &         \\
\nuc{Yb}{168} ($0\nu\rm KL$) & \nuc{Er}{168} ($0^+_1 1217.2)$   &   1137.4    & 1.81   &   1.03  &      \baseT{>7.1}{16}            \\
											 &  &   124.8    & 0.78   &   0.81  &            \\
\nuc{Yb}{176} (\bb{2}) & \nuc{Hf}{176} ($2^+_1 88.3)$   &   88.3    & 0.015   &   0.80  &   \baseT{>4.7}{16} (*)               \\

\hline
\end{tabular}
\medskip
\caption{\label{tab:HLimits} Lower half-life limits on investigated decay modes of Yb isotopes. Column 3-5 show the \glines\ used in the fit together with their detection efficiency and resolution. In case of multiple \glines, a combined fit is used for the limit setting.
(*) No discovery is possible for these decay modes in this search due to overlapping background \glines. The given limits are valid nevertheless.
}
\end{center}
\end{table*}

A prior probability is assigned to each free parameter. The prior distribution for the inverse half-life $(T_{1/2})^{-1}$ and the linear background parameters is flat. For energy resolution, peak position and detection efficiencies, the priors are Gaussian distributions centered around the mean values of these parameters and a width of the parameter uncertainty. 
This naturally includes the systematic uncertainty into the fit result.

The uncertainty of the peak positions are set to 0.1~keV.    
The energy scale and resolution is obtained from standard calibration spectra with an estimated uncertainty of 5\%.
The full energy peak detection efficiencies are determined with MaGe Monte-Carlo simulations \cite{bos11} with an estimated uncertainty of 10\%. 
Systematic uncertainties on the measured sample mass and the isotopic fraction in the sample are small with respect to the uncertainty of the detection efficiency and are neglected. 

Background \glines\ above 1\% emission probability from the \nuc{U}{238} and \nuc{Th}{232} decay chains are included in the fit windows. 
Also included are \glines\ from the \nuc{Lu}{176} and \nuc{Yb}{169} isotopes in the target powder where applicable. 
For these searches this affects only the low energy region through
63.1~keV (4.3\%, \nuc{Yb}{169}), 
63.2~keV (3.6\%,  \nuc{Lu}{176}), 
63.3~keV (2.7\%, \nuc{Th}{234}), 
64.9~keV (1.2\%,  \nuc{Lu}{176}), 
84.4~keV (1.2\%, \nuc{Th}{228}), 
88.3~keV (14.5\%,\nuc{Lu}{176}), 
92.4~keV (2.1\%, \nuc{Th}{234}), 
92.8~keV (2.1\%, \nuc{Th}{234}), 
93.6~keV (2.6\%, \nuc{Yb}{169}), 
99.5~keV (1.3\%, \nuc{Ac}{228}), and
129.1~keV (2.4\%, \nuc{Ac}{228}) \grays,
with emission probabilitiy and isotope in parenthesis. 
The 88.3~keV \gline\ from \nuc{Lu}{176} originates from the same final state in \nuc{Hf}{176} as the \nuc{Yb}{176} \bb{2} $2^+_1$ transition, making this decay mode signal indistinguishable from background. Hence, no discovery is possible in this case. Nevertheless, peak counts and thus a half-life can be constrained with a limit. 
Similarly, for the \nuc{Yb}{173} $\alpha$-decay, the 64.6~keV signal \gline\ is too close to background \glines\ to be resolved. Also here only a half-life exclusion below a certain value is possible.  


The posterior probability distribution is calculated from the likelihood and prior probabilities with BAT. It is then marginalized for $(T_{1/2})^{-1}$. 
The best fit values for all decay modes is consistent with zero signal counts and the 0.9 quantile of the distribution is used to set 90\% credibility limits. 

The results are shown for all investigated decay modes in \tab \ref{tab:HLimits}. Also shown are the \gline\ energies, the full energy peak detection efficiency and the resolution for the \glines\ used in the analyses. 
For the  \nuc{Yb}{168} \bbe{2} $0^+_1$, $0\nu{\rm KK}\ 2^+_1$, $0\nu{\rm KK}\ 2^+_2$, and $0\nu{\rm KL}\ 0^+_1$ decay modes, multiple \glines\ are used, each having its own fit window, likelihood, and free parameters but sharing the same half-life parameter in \eq \ref{eq:HLtoCounts}.

The half-life limits are strongly dependent on the isotopic abundance and the detection efficiency. 
For $\alpha$-decays into excited states, the lower limits range from \baseT{6}{14}~yr to \baseT{2}{16}~yr.
For double beta decay modes, half-lives up to \baseT{8}{16}~yr could be excluded.

\section{Discussion and Conclusion}

A general search for rare nuclear decays was performed in ytterbium isotopes using an ultra-low background HPGe $\gamma$-spectroscopy setup and a 194.7~g Yb$_2$(C$_2$O$_4$)$_3$ powder sample. 
The search included alpha decays, double beta decays and double electron captures into excited states of the daughter isotopes. No signal was found and 90\% credibility limits were set using a Bayesian analysis. These results are the first constraints on the $\alpha$-decay modes. Existing constraints on double beta decay and double electron capture mode could be confirmed and partially improved. The achieved sensitivity, however, is still far away from theoretical predictions (\tab \ref{tab:isotopes}).

General improvements for this measurement can be achieved through background reduction and increased mass and measurement time.  
The natural radionuclide content of the sample was already very low. 
No significant contribution from the natural decay chains \nuc{U}{235}, \nuc{U}{238}, and \nuc{Th}{232}, nor from \nuc{K}{40}, \nuc{Co}{60}, and \nuc{Cs}{27} were detected.
Some radioactivity of \nuc{Yb}{169}, \nuc{Yb}{175} and \nuc{Lu}{176} was observed. However, it did not strongly interfere with the measurement.
The Yb$_2$(C$_2$O$_4$)$_3$ sample in this work contained a significantly lower contamination of \nuc{Lu}{176} (15~mBq/kg) compared to the Yb$_2$O$_3$ sample used in a previous search \cite{Belli19_Yb168} (420~mBq/kg). This allowed for a lower background in searches of decay modes with \gray\ emissions below 300~keV. 
The precise control of the \nuc{Lu}{176} contamination is also the only way to probe the \nuc{Yb}{176} \bb{2} $2^+_1$ transition which shares the same de-excitation \gray.
Further purification can be achieved through liquid-liquid extraction of inorganic ytterbium compounds followed by their transformation to metallo-organic Yb$_2$(C$_2$O$_4$)$_3$. A factor of 10 reduction of \nuc{K}{40}, U/Th-chain backgrounds has been demonstrated in CeO$_2$, Gd$_2$O$_3$, and Nd$_2$O$_3$, as described in \cite{Polischuk13}.
Mitigation of cosmogenically produced short-lived \nuc{Yb}{169} and \nuc{Yb}{175} isotopes can be achieved by underground storage for multiple half-lives of a few months.
 The  measurement time of 11~d could be significantly improved in future measurements. A 3~yr measurement would increase the sensitivity by about an order of magnitude. 

To approach the predicted half-lives, a drastic improvement of detection efficiencies is necessary. Realistically, this can only be achieved by applying the ``source = detector'' approach, where target isotopes are embedded into the detector material. 
Crystals, such as Yb$_2$SiO$_5$, operated as scintillators, as bolometers, or as a combination of both could be used. 
This would especially benefit the low energy region where multiple orders of magnitude improvement in detection efficiency can be achieved. Alphas would be detected directly, making ground state transitions accessible. Pulse-shape discrimination of alpha and electron recoil interaction would allow a practically background-free search for $\alpha$-decays modes.  

Ultimately, enrichment of specific ytterbium isotopes could gain another one or two orders of magnitude improvement in sensitivity for a focused search in a certain ytterbium isotope. However, this procedure is expensive (level of \$1000/mg) and only feasible for relatively small target masses. 

\begin{acknowledgements}

These results have been obtained as a by-product of internal radioactive contamination studies for Yb$_2$(C$_2$O$_4$)$_3$ compounds in connection with the 
LENS (Low Energy Neutrino Search) experiment \cite{lens}.
\end{acknowledgements}

\bibliographystyle{spphys}       


\end{document}

%% file: commands.tex
\newcommand{\nuc}[2]{$^{#2}\rm #1$}

\newcommand{\bb}[1]{$\rm #1\nu \beta \beta$}
\newcommand{\bbm}[1]{$\rm #1\nu \beta^- \beta^-$}
\newcommand{\bbp}[1]{$\rm #1\nu \beta^+ \beta^+$}
\newcommand{\bbe}[1]{$\rm #1\nu \epsilon \epsilon$}
\newcommand{\bbep}[1]{$\rm #1\nu \rm EC \beta^+$}

\newcommand{\pic}[5]{
       \begin{figure}[ht]
       \begin{center}
       \includegraphics[width=#2\textwidth, keepaspectratio, #3]{#1}
       \end{center}
       \caption{#5}
       \label{#4}
       \end{figure}
}

\newcommand{\apic}[5]{
       \begin{figure}[H]
       \begin{center}
       \includegraphics[width=#2\textwidth, keepaspectratio, #3]{#1}
       \end{center}
       \caption{#5}
       \label{#4}
       \end{figure}
}

\newcommand{\sapic}[5]{
       \begin{figure}[P]
       \begin{center}
       \includegraphics[width=#2\textwidth, keepaspectratio, #3]{#1}
       \end{center}
       \caption{#5}
       \label{#4}
       \end{figure}
}

\newcommand{\picwrap}[9]{
       \begin{wrapfigure}{#5}{#6}
       \vspace{#7}
       \begin{center}
       \includegraphics[width=#2\textwidth, keepaspectratio, #3]{#1}
       \end{center}
       \caption{#9}
       \label{#4}
       \vspace{#8}
       \end{wrapfigure}
}

\newcommand{\baseT}[2]{\mbox{$#1\times10^{#2}$}}
\newcommand{\baseTsolo}[1]{$10^{#1}$}
\newcommand{\THL}{$T_{\nicefrac{1}{2}}$}

\newcommand{\UBI}{$\rm cts/(kg \cdot yr \cdot keV)$}

\newcommand{\Uflux}{$\rm m^{-2} s^{-1}$}
\newcommand{\Ucpd}{$\rm cts/(kg \cdot d)$}
\newcommand{\Uexpo}{$\rm kg \cdot d$}

\newcommand{\Qbb}{$\rm Q_{\beta\beta}\ $}

\newcommand{\validate}{\textcolor{blue}{\textit{(validate!!!)}}}

\newcommand{\improve}{\textcolor{blue}{\textit{(improve!!!)}}}

\newcommand{\missing}[1]{\textcolor{red}{\textbf{...!!!...} #1}\ }

\newcommand{\missref}{\textcolor{red}{[reference!!!]}\ }

\newcommand{\quanta}{\textcolor{red}{\textit{(quantitativ?) }}}

\newcommand{\misscite}{\textcolor{red}{[citation!!!]}}

\newcommand{\PC}{$N_{\rm peak}$}
\newcommand{\BIC}{$N_{\rm BI}$}
\newcommand{\PAPR}{$R_{\rm p/>p}$}

\newcommand{\PCR}{$R_{\rm peak}$}


\newcommand{\gline}{$\gamma$-line}
\newcommand{\glines}{$\gamma$-lines}
\newcommand{\gray}{$\gamma$-ray}
\newcommand{\grays}{$\gamma$-rays}


\newcommand{\tab}{Tab.~}
\newcommand{\eq}{Eq.~}
\newcommand{\fig}{Fig.~}
\renewcommand{\sec}{Sec.~}
\newcommand{\chap}{Chap.~}

 \newcommand{\fn}{\iffalse \fi} 
 \newcommand{\tx}{\iffalse \fi} 
 \newcommand{\txe}{\iffalse \fi} 
 \newcommand{\sr}{\iffalse \fi} 